# Elastic anomalies associated with domain switching in BaTiO$_3$ single crystals under *in-situ* electrical cycling

D. Pesquera,[1,*] B. Casals,[1] J. E. Thompson,[2] G. F. Nataf,[2] X. Moya,[2] and M. A. Carpenter[1]

[1]*Department of Earth Sciences, University of Cambridge, Downing Street, Cambridge CB2 3EQ, United Kingdom*

[2]*Department of Materials Science, University of Cambridge, 27 Charles Babbage Road, Cambridge CB3 0FS, United Kingdom*

\* dpesquera@cantab.net

The elastic response of BaTiO$_3$ single crystals during electric field cycling at room temperature has been studied using *in-situ* Resonant Ultrasound Spectroscopy (RUS), which allows monitoring of both the elastic and anelastic changes caused by ferroelectric polarization switching. We find that the first ferroelectric switching of a virgin single crystal is dominated by ferroelastic 90° switching. In subsequent ferroelectric switching, ferroelastic switching is reduced by domain pinning and by the predominance of 180° ferroelectric domains, as confirmed by polarized light microscopy. RUS under *in-situ* electric field therefore demonstrates to be an effective technique for the investigation of electromechanical coupling in ferroelectrics.

The process of polarization switching in ferroelectrics is generally not homogeneous and involves nucleation and growth of domains [1,2]. In prototypical tetragonal ferroelectrics, such as BaTiO$_3$ at room temperature, two kinds of ferroelectric domains can nucleate [3]: 180° domains with polarizations antiparallel to each other, which minimize depolarization fields, and 90° domains with polarizations orthogonal to each other, which minimize strain via the formation of twins. During electrical poling, local strains are generated when 90° domains switch, but not when 180° domains switch. Limiting the mobility of domain walls by introducing specific point defects and domain engineering has proven an effective strategy to enhance piezoelectricity in BaTiO$_3$ [4,5]. Large strains are desirable, e.g., to achieve giant magnetoelectric effects in multiferroic composites [6–10], and polarization switching without domain propagation is preferred for ferroelectric memory devices [11,12]. Separately, domain wall pinning and domain jamming may cause fatigue effects in ferroelectrics, hampering polarization switching upon repeated voltage cycling, ultimately limiting the performance of devices [13,14]. Understanding the role of domain structure on ferroelectric switching is thus of high technological relevance for the optimization of devices based on ferroelectric crystals. BaTiO$_3$ has been the subject of many such studies for more than fifty years [15–23]: some focused on the strain coupling with polarization switching [20,22,23], and some on the microstructure evolution with electric field [15–19,21]. However, combined experiments showing the link between both aspects are lacking.

RUS is a convenient technique for determining the elastic constants of solids. Since the elastic constants are second derivatives of the free energy, their evolution as a function of external stimuli provides a highly sensitive indicator of strain coupling with the driving order parameter of a phase transition [24]. Moreover, the small dynamic stress that is applied to the sample during RUS is well suited for exploring relaxation mechanisms in the elastic regime, and for studying dissipation processes due to microstructural dynamics as associated with e.g. twin-wall motion [25,26]. RUS has also proved to be an effective tool for identifying thermally driven energy dissipation associated with the existence of polar nanoregions in BaTiO$_3$ [27] and relaxors [28,29], but the elastic and anelastic responses under *in-situ* electrical cycling remain unexplored.

Here we report, for the first time, RUS measurements under *in-situ* electric field in the prototypical ferroelectric BaTiO$_3$. We show that this technique allows for the identification of strain and energy dissipation caused by ferroelectric switching processes. We find that both strain and dissipation are large upon the first electrical poling, but gradually disappear after a few electrical switches, suggesting a reduced contribution of ferroelastic switching, and a freezing of domain dynamics, under sustained operation. Complementary *in-situ* studies of domain distribution using polarized light microscopy reveal the evolution from initially predominant 90° switching via twin walls motion, to localized 180° switching, limited by domain pinning around defects and locally strained regions. These observations are in agreement with the results obtained from RUS, thus demonstrating the potential of RUS to explore ferroelectric switching mechanisms.

The RUS data and optical images reported here were taken on two different pieces (labelled samples A and B) of an optically transparent electrically virgin BaTiO$_3$ single crystal (MaTecK), which was cut using a fine annular diamond saw. Each piece had approximate dimensions of 3 mm x 3 mm x 0.5 mm, with edges parallel to <100> crystallographic axes. Initial optical examination using a Zeiss Axioplan 1 optical microscope in transmission mode under crossed polarizers, and a digital camera, revealed that both samples presented a similar configuration of micron-size domains with boundaries parallel to [100], evidencing a mixed microstructure consisting of *a* domains (with polarization vectors along [100] or [010]) and *c* domains (with polarization vectors along [001]).

For RUS measurements under electric field, sample A was held by two opposite corners between two piezoelectric transducers. 20 μm-thick copper wires were attached to the two large surfaces parallel to (001) using RS-PRO silver conducting paint as electrodes. DC voltage was applied to the sample via the core and the shield of a coaxial cable connected to a Keithley 2260B power supply. At each value of applied voltage, mechanical resonances of the crystal in the frequency range 400-1300 kHz were measured using a Lock-in amplifier (Stanford, model SR830) [30].

For optical microscopy under electric field, ~200 nm thick transparent $InTiO_3$ electrodes were deposited at a base pressure $< 10^{-8}$ mbar using magnetron sputtering (20 W, 2.5 Pa Ar) on both (001) surfaces of sample B. During the imaging, an increasing voltage was applied across the sample at a rate of 0.3 V/s using a Keithley 6487 picoammeter, which also allowed the displacement current in the sample to be recorded.

In RUS experiments on single crystals, the frequencies of individual mechanical resonances are determined by combinations of elastic constants, with a predominant contribution from shear modes. For the first two electrical cycles, Fig.1(a,b) shows the evolution of a resonance peak in sample A that does not interfere with adjacent peaks. Other peaks were found to display the same qualitative evolution with applied voltage. The results of fitting this peak with an asymmetric Lorentzian function are given in Fig.1(c,d), as (i) values of the squared resonance frequency, $f^2$, which scale with the elastic constants, and (ii) the inverse mechanical quality factor $Q^{-1} = \Delta f/f$, which is a measure of acoustic attenuation ($\Delta f$ is the peak full width at half maximum). A pronounced softening of the elastic constants with increasing electric fields is evident upon the first poling of the crystal. The onset of this softening occurs simultaneously with a sharp peak in the acoustic dissipation at the coercive field $E_c \approx 0.4$ kV/cm. On further increasing field, the crystal stiffens (increase in $f^2$) and a wider peak in the dissipation is observed around $2E_c$. When approaching the highest applied field (2 kV/cm), the initial resonance frequency and attenuation values are recovered, and these values are preserved after removing the voltage. When reversing the electric field, a smaller elastic softening and a smaller increase in the acoustic loss are evidenced, as revealed by a minimum in $f^2$ and a maximum in $Q^{-1}$ around $-E_c$, respectively.

The butterfly-shaped curve observed in the initial cycle is modified in the second electrical cycle (Fig.1d), where a softening at $+E_c$ is only partially compensated when applying a negative field. In this second cycle, acoustic losses remained constant and below 0.5 %. Further cycling the electric field yielded smaller changes in the elastic constants (cyan points in Fig.1d) and negligible variations in the attenuation.

While the DC applied voltage can move both 90° and 180° twin walls because both are ferroelectric, the small dynamic stress applied during RUS measurements only swings 90° walls, as they are ferroelastic (180° walls are unaffected as the shear strain is identical in the neighboring domains). Therefore, RUS only probes the elastic and anelastic changes associated with 90° twin walls. In this light, we can interpret the results presented above as an evolution of the twin wall population from an initial state dominated by 90° walls (which can be either *a-a* or *a-c*) to a final state with predominantly 180° walls (*c-c*). The initial poling causes the motion of 90° twin walls, hence inducing an elastic softening. The two peaks observed in the acoustic loss indicate the occurrence of two domain switching processes taking place during the first cycling. When the twin walls reach new equilibrium positions, or *a* domains transform into *c* domains, the dissipation is reduced and the crystal stiffens. In the process of polarization reversal, a single broad softening and loss peak indicate ferroelastic switching around $-E_c$. However, on successive cycling, RUS spectra evidence frozen dynamics and minor changes in the elastic constants, pointing to increased domain pinning and/or predominant non-ferroelastic 180° switching. We note that wall mobility is only recovered after annealing the crystal above $T_c$, as shown in the Supplementary Information.

In order to understand the microscopic mechanism behind the observed evolution of the elastic constants and acoustic dissipation, we imaged sample B while sweeping the applied voltage (Fig.2a and Supplementary Videos), and simultaneously measuring the displacement current (Fig.2b,c). 90° domains in the virgin crystal were clearly visible due to the birefringence contrast between them: dark regions showing no birefringence contrast correspond to *c* domains, while bright regions correspond to *a* domains. The crossed polarizers were adjusted at 45 degrees from the main crystallographic axes to maximize the contrast. In the initial state (step 1), all domain walls were aligned with the [100] axis. Regions with large *a* domains and *c* domains can be seen at the bottom of the image, while a pattern of fine *a-c* domains is seen at the top. Such domain width variations across the crystal were also observed in sample A, and may be evidence of an inhomogeneous distribution of defects [31].

When an electric field was applied during initial poling, the large *a* domain was swept away (step 2) and merged completely with the bottom *c* domain at a field $E_c \approx 0.4$ kV/cm. Further increasing the voltage above $2E_c$ caused a widening of the top *a-c* domains and jerky movements of birefringent regions away from the *c* macrodomain (step 3, Supplemental Video 1). More subtly, a number of bubble-like domains nucleated in the *c* macrodomain, suggesting some minority 180° switching [32], that became more evident in the following electrical cycles. At the highest field, no color contrast was observed, as expected for a single *c* domain state, though weak contrast from the *a-c* pattern could still be observed at the top of the image (panel 4). This may correspond to pinned domains, likely localized at the surface of the crystal [33].

The displacement current measurements (Fig.2b) confirm that initial poling of the sample is dominated by two polarization switching events, as revealed by a large current peak at $E_c$, followed by a number of smaller current jerks around $2E_c$. Both processes can be correlated with the ferroelastic switching events observed in the optical images, corresponding to displacement of domain walls due to the 90° switching from a multidomain *a-c* structure to a nominally single *c* domain (See Supplemental Video 1 for the complete poling process).

The process of polarization reversal is illustrated in steps 5-7 of Fig. 2a (shown fully in Supplemental Video 2). An increase in the birefringence, with simultaneous multiple peaks in the switching current (Fig.2c), indicates nucleation of *a-c* domains in their previous locations. This is followed by the propagation of a corrugated structure towards the *a-c* pattern (step 6), occurring at $-E_c$ and producing a large peak in the current. This kind of structure is typically observed in the switching of antiparallel domains [34,35], and here provides evidence of the nucleation and propagation of 180° domain walls. During propagation, the front is slowed down by defects that act as pinning centers. At higher fields the contrast becomes uniform, except for the remaining lines that arise from the *a-c* surface pattern, and increased texturing in the whole area (step 7). On reversing the voltage, the nucleation of *a-c* domains follows mostly the previous behaviour, although the 180° switching in the *c* macro-domain occurs via nucleation of bubble-like domains with no propagation (see Supplemental Video 3), favorably at the location of the defects that acted as nucleation centers (step 8). The alignment of these domains parallel to pre-existing domain walls suggests a nucleation mechanism via skyrmion domain ejection, as proposed by Dawber *et al.* [36]. Nucleation occurs at the same time as the large increase in the displacement current (Fig.2c). Traces of these bubble domains could still be seen at the highest applied field (step 9), which may indicate the presence of localized regions of remanent strain.

To summarize the above observations, while in the first poling most switching events occur via motion of 90° domain walls, evidencing their lower energy with respect to 180° walls [37], in the electrical cycle that follows the first poling, 180° domain switching becomes the predominant source for polarization reversal, although a significant contribution from ferroelastic switching is still noticeable in some regions of the sample. These results are consistent with previous observations in which 180° domain nucleation in poled crystals was found to be easier than 90° domain nucleation, due to strain constraints [34], and with the observed reduction of acoustic emission signals in BaTiO$_3$ crystals after the first electrical cycle [23], suggesting that 90° domain walls are more prone to pinning. The total polarization that switches reversibly in the electrical cycling shown in the first cycle is 20.6 µC/cm$^2$ (Fig.1b,c).

On subsequent cycling, however, the switched polarization gradually drops, as evidenced by the decreasing amplitude of the displacement current peak observed in the *I-V* curves (Fig. 3a). In the second and third cycle the switched polarization is reduced by 44 and 76%, respectively (inset in Fig. 3a). A significantly reduced domain motion was also identified in the optical images (see Supplemental Video 4). When increasing the maximum electric field to 8 kV/cm, small current peaks still occur at ±$E_c$, but much larger and broader peaks are observed around ±4 kV/cm (Fig.3b). A large increase in the leakage current is also observed –likely caused by sample damage at high voltages- which prevents a reliable estimation of the switched polarization. During this higher-field electrical cycling, no clear signature of domain switching is identified in the images (see Supplemental Video 5), which show a uniform structure of islets with edges partially aligned to the main in-plane crystallographic axes, and some regions of stripes parallel to [100] showing weak birefringence (inset in Fig.3b). These observations suggest a structure dominated by pinned *c* domains, which switch uniformly at high fields (with no propagation) and some remaining areas of *a-c* domains, also pinned at the locations reached after the initial cycling of the crystal, which still switch at low field. Such observations point towards notable fatigue effects in our BaTiO$_3$ samples, likely induced by agglomeration of defects at domain walls during electrical cycling [38–40], or electric field induced cracking [41], which lead to domain wall pinning and increased coercive fields. We note that the fatigue effects that we observe in our samples may be dependent on measurement conditions (e.g. field frequency and amplitude) and sample dimensions (thicker crystals may be less affected by surface pinning and develop further ferroelastic switching [23]).

In summary, we have shown that RUS allows *in-situ* monitoring of ferroelastic switching processes during electrical cycling of ferroelectric crystals. Large changes in elastic properties and acoustic loss accompany the redistribution of domains (and strains) in the crystal during the first poling, and much smaller changes occur in subsequent electrical cycles. Polarized light microscopy confirms the occurrence of inhomogeneous strain due to motion of ferroelastic domains walls during the first poling. 180º polarization switching, which does not contribute to the elastic changes, becomes predominant after a few cycles, and pinning due to defects and localized strains cause enhancement of the ferroelectric coercivity.

RUS while *in-situ* cycling electric field may open up new possibilities in, for example: (i) studies of engineered ferroelectric samples [12] where domain structures are prepared to favor a defined polarization switching process; (ii) exploration of the dependence of applied field orientation on the switching mechanism [22], or on the elastic behavior near phase transitions [42,43]; (iii) identification of the elastic signatures of different possible switching pathways for lower symmetry phases [44,45]; and (iv) investigations of electric-field response of polar regions near ferroelastic domain walls in non-polar materials [46–49].

**Supplementary Material**

See supplementary material for the RUS measurements on sample A after annealing, and description of supplementary videos taken on sample B.

**Acknowledgments**

This work was funded by EPSRC grant no. EP/P024904/1, and ERC Starting Grant No. 680032, which are gratefully acknowledged. RUS facilities in Cambridge were established through grants from the Natural Environment Research Council (grant nos. NE/B505738/1, NE/F017081/1) and the Engineering and Physical Sciences Research Council (grant no. EP/I036079/1) to MAC. GFN would like to thank the Royal Commission for the Exhibition of 1851 for the award of a Research Fellowship. JET and GFN are grateful to Nadia Stelmashenko for useful discussions. X. M. is grateful for support from the Royal Society.

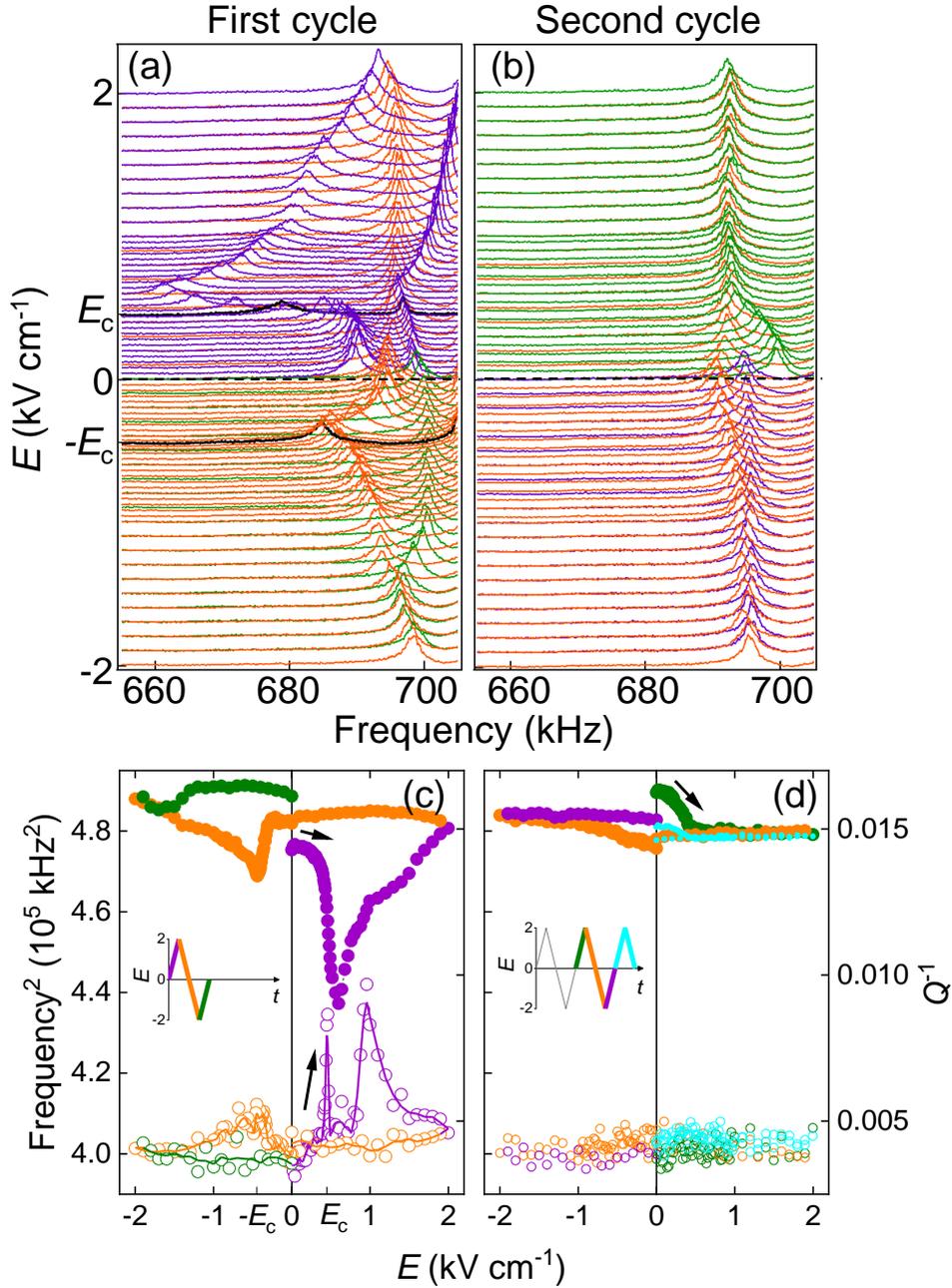

FIG. 1. Detail of the electric-field evolution of RUS spectra during (a) first, and (b) second electrical cycles in an initially electrically virgin BaTiO$_3$ crystal (sample A). Each spectrum has been offset up the $y$-axis in proportion to the voltage applied to the sample during the acquisition of the spectrum. Spectra colors indicate the field ramp step as shown in (c,d); black spectra in (a) correspond to spectra taken at $\pm E_c$ on increasing field. Dependence with applied voltage of $f^{\,2}$ (left axis, closed circles) and $Q^{-1}$ (right axis, open circles) from fits to the resonance peaks in (a,b) are shown in (c,d). Insets show the applied field sequence.

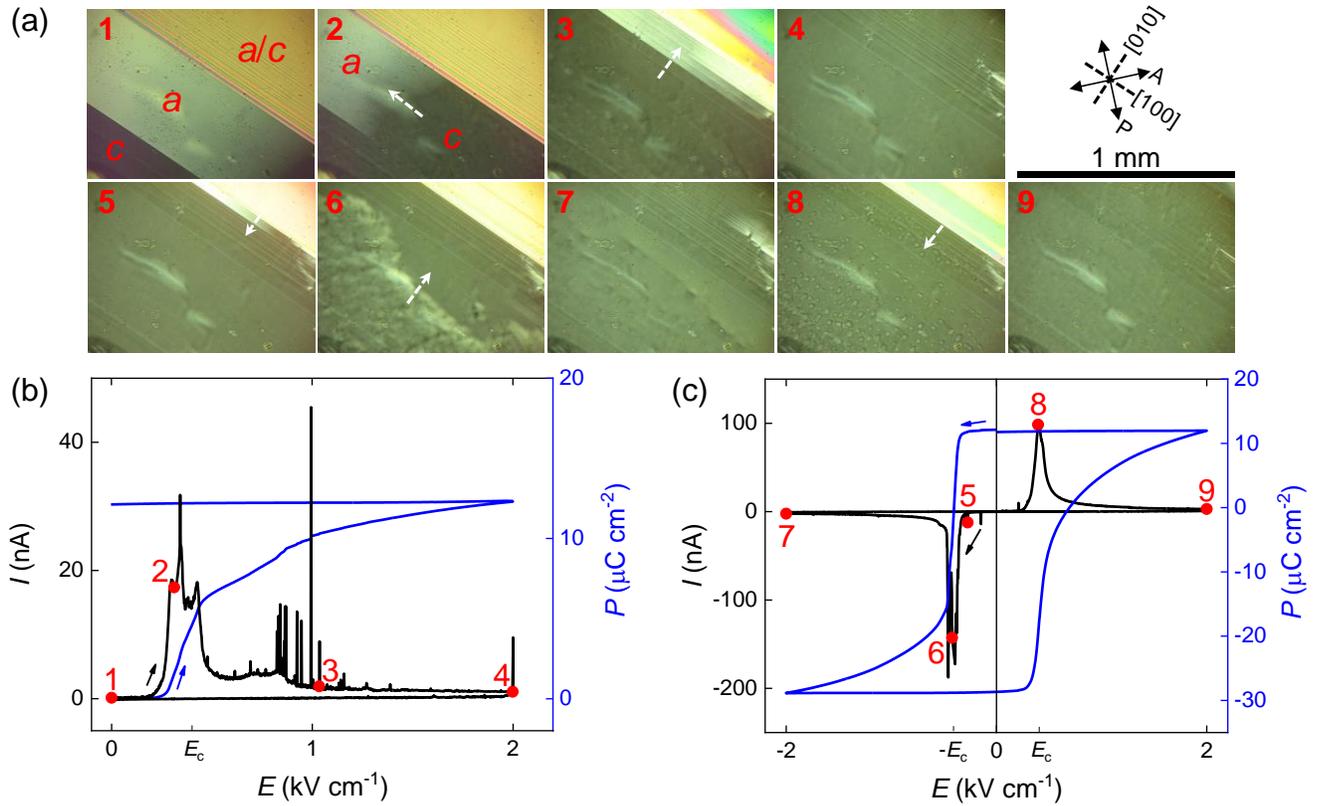

FIG. 2. (a) Snapshots of a region of sample B during electrical cycling, taken at steps indicated in (b,c), using light transmitted along [001] of the crystal and crossed polarizers, both at 45 degrees from the main in-plane <100> axes (see sketch at top right corner). The voltage was applied along [001] through transparent InTiO$_3$ electrodes deposited on opposite sides of the crystal. Labels in panels 1 and 2 indicate the orientations of BaTiO$_3$ domains, with $a$ domains having polarization along [010], and $c$ domains having polarization along [001]. White dashed arrows indicate the observed domain-wall propagation during electrical cycling (see Supplementary Videos for the complete evolution of the domain structure under field). (b) Electric–field ($E$) dependence of displacement current $I$ (black, left axis), and corresponding polarization $P$ (blue, right axis) during the initial poling. (c) $I(E)$ (left axis) and $P(E)$ (right axis) for a complete electrical cycle after the first poling.

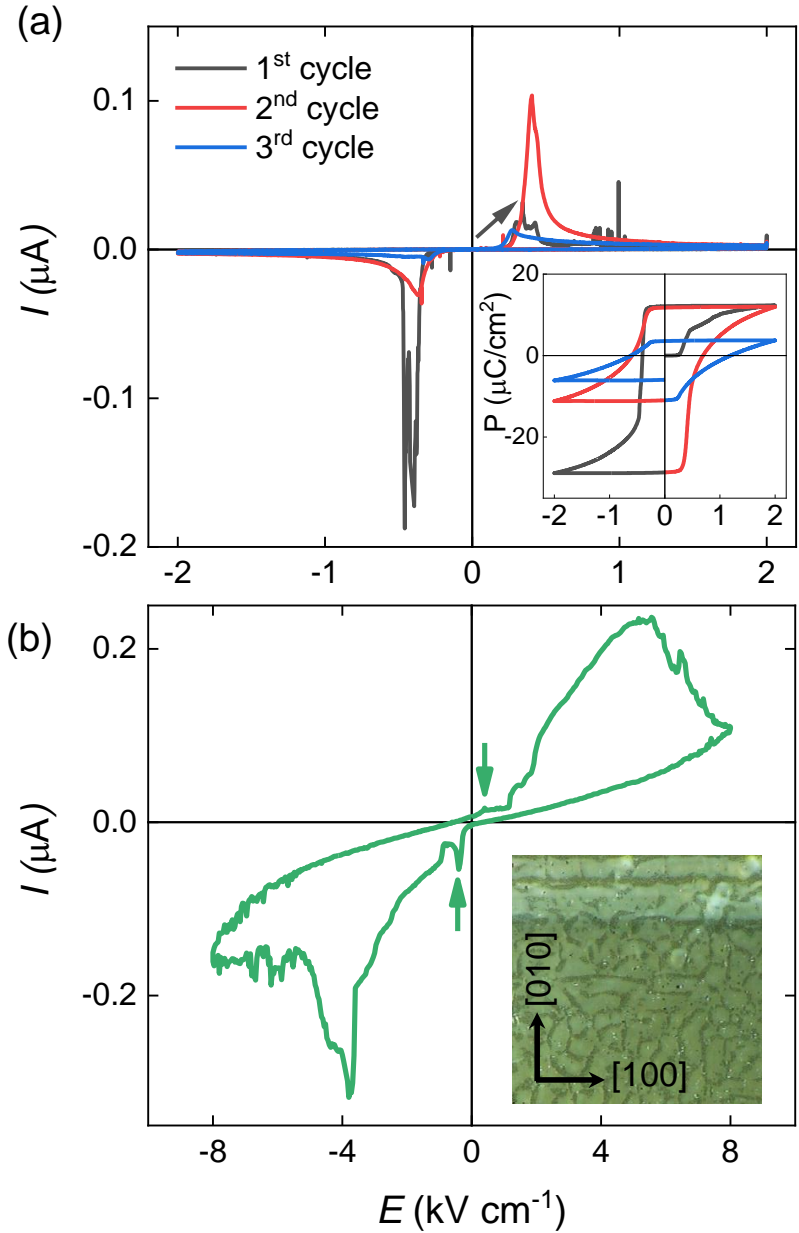

FIG. 3. (a) Displacement current during electrical cycling of sample B with a maximum electric field of 2 kV/cm. Inset: Polarization loops obtained from the integration of displacement current (b) Displacement current versus electric field, when increasing the maximum field to 8 kV/cm; arrows indicate small current peaks observed at $\pm E_c$. The inset image shows a 500 x 500 μm region of the sample after repeated electrical cycling up to ±8 kV/cm.